\documentclass[journal=langd5,manuscript=article,layout=twocolumn]{achemso}
\setkeys{acs}{etalmode=truncate,maxauthors=3}

\makeatletter
\let\l@addto@macro\relax
\makeatother
\usepackage[fontsize=10pt]{scrextend}

\usepackage{graphicx}
\usepackage{textgreek}
\usepackage{tabularx}
\usepackage{array}

\usepackage{amssymb}
\usepackage{amsmath}

\usepackage{textcomp}
\usepackage{gensymb}

\usepackage{textcomp}

\usepackage{xr}

\usepackage{xcolor} 
\mciteErrorOnUnknownfalse 

\usepackage{acro}			

\DeclareAcronym{AFM}{short = AFM, long = atomic force microscopy}
\DeclareAcronym{CD}{short = CD, long = circular dichroism}
\DeclareAcronym{DIC}{short = DIC, long = differential interference contrast}
\DeclareAcronym{DLS}{short = DLS, long = dynamic light scattering}
\DeclareAcronym{PSF}{short = PSF, long = point spread function}
\DeclareAcronym{qDIC}{short = qDIC, long = quantitative differential interference contrast}
\DeclareAcronym{RCF}{short = RCF, long = relative centrifugal force}
\DeclareAcronym{TEM}{short = TEM, long = transmission electron microscopy}
\DeclareAcronym{aptes}{short = APTES, long = 3-aminopropyl triethoxysilane}
\DeclareAcronym{bf}{short = BF, long = bright-field}
\DeclareAcronym{bfp}{short = BFP, long = back focal plane}
\DeclareAcronym{ccd}{short = CCD, long = charge-coupled device}
\DeclareAcronym{df}{short = DF, long = dark-field}
\DeclareAcronym{fwhm}{short = FWHM, long = full width at half maximum}
\DeclareAcronym{led}{short = LED, long = light-emitting diode}
\DeclareAcronym{lspr}{short = LSPR, long = localized surface plasmon resonance}
\DeclareAcronym{na}{short = NA, long = numerical aperture}
\DeclareAcronym{np}{short = NP, long = nanoparticle}
\DeclareAcronym{pw}{short = PW, long = plane wave}
\DeclareAcronym{si}{short = SI, long = supporting information}

\DeclareAcronym{slb}{short = SLB, long = supported lipid bilayer}
\DeclareAcronym{DOPC}{short = DOPC, long = {1,2-dioleoyl-sn-glycero-3-phosphocholine}}
\DeclareAcronym{dcpc}{short = DC\textsubscript{15}PC, long={1,2-dipentadecanoyl-sn-glycero-3-phosphocholine}}
\DeclareAcronym{DOPE}{short = DOPE, long = {1,2-dioleoyl-sn-glycero-3-phosphoethanolamine}}
\DeclareAcronym{SM}{short = ESM, long = chicken egg sphingomyelin}

\DeclareAcronym{PBS}{short = PBS, long = phosphate-buffered saline}
\DeclareAcronym{GUV}{short = GUV, long = giant unilamellar vesicle}
\DeclareAcronym{SUV}{short = SUV, long = small unilamellar vesicle}

\DeclareAcronym{er}{short = ER, long = endoplasmic reticulum}

\newcommand{\Eq}[1]{Eq.(\ref{#1})}
\newcommand{\Fig}[1]{Fig.\,\ref{#1}}
\newcommand{\Sec}[1]{Sec.\,\ref{#1}}
\newcommand{\Onlinecite}[1]{Ref.\nocite{#1}\citenum{#1}}

\newcommand{\be}{\begin{equation}}
	\newcommand{\ee}{\end{equation}}

\newcommand{\So}{S\textsubscript{o}}
\newcommand{\Ld}{L\textsubscript{d}}
\newcommand{\Lo}{L\textsubscript{o}}

\newcommand{\nor}{\ensuremath{n_{\rm o}}}
\newcommand{\nex}{\ensuremath{n_{\rm e}}}

\newcommand{\Ic}{\ensuremath{I_{\rm c}}}

\newcommand{\br}{\ensuremath{\mathbf{r}}}

\DeclareMathOperator{\sech}{sech}

\title{Understanding the mechanisms of supported lipid membrane reshaping into tubular networks using quantitative DIC microscopy}
\author{David Regan}
\affiliation{School of Physics and Astronomy, Cardiff University, The Parade, Cardiff CF24 3AA, UK}
\author{Paola Borri}
\affiliation{School of Biosciences, Cardiff University, Museum Avenue, Cardiff CF10 3AX, UK}
\author{Wolfgang Langbein}
\affiliation{School of Physics and Astronomy, Cardiff University, The Parade, Cardiff CF24 3AA, UK}

\begin{document}


\begin{abstract}
	Biological membranes are known to form various structural motifs, from lipid bilayers to tubular filaments and networks facilitating e.g. adhesion and cell-cell communication. To understand the biophysical processes underpinning lipid-lipid interactions in these systems, synthetic membrane models are crucial. Here, we demonstrate the formation of tubular networks from supported lipid membranes of controlled lipid composition on glass. We quantify tube radii using \ac{qDIC} and propose a biophysical mechanism for the formation of these structures, regulated by surface tension and lipid exchange with connected supported membranes. Two lipid types are investigated, namely \ac{DOPC} and \ac{dcpc}, exhibiting a liquid disordered and a solid ordered phase at room temperature, respectively. Tube formation is studied versus temperature, revealing bilamellar layers retracting and folding into tubes upon \ac{dcpc} lipids transitioning from liquid to solid phase, which is explained by lipid transfer from bilamellar to unilamellar layers. This study introduces a novel model system for bilayer tubes, allowing to elucidate the biophysics of lipid-lipid interactions governing lipid membrane reshaping into tubular structures, important for our understanding of biological membrane filaments.               
\end{abstract}

\section{Introduction}

In our endeavour to understand the biophysical properties of cellular membranes, few innovations have had as significant an impact as the development of synthetic membrane model systems.\cite{NisticoAMT23} By allowing the formation of lipid bilayer membranes in controlled geometries with defined compositions, membrane model systems enabled probing aspects of intrinsic membrane behaviour in isolation, yielding key insights into cellular activity, including membrane fusion \cite{MarsdenCSR11}, membrane fluctuations\cite{MonzelJPD16}, the behaviour of membrane proteins \cite{JorgensenEBJ17}, and the proposed segregation of lipids into phase separated raft domains.\cite{DimovaACIS14}

The two most commonly used membrane model systems are synthetic unilamellar vesicles\cite{NisticoAMT23} and \acfp{slb}\cite{NisticoAMT23}, which act as analogues for specific types of cellular structures. \Acp{SUV}, with diameters of tens to hundreds of nanometres, are structurally analogous to biological vesicles, while larger \acp{GUV}, which are tens of microns in size, and planar \acp{slb} reconstitute the weak curvature of cell plasma membranes. Notably one type of membrane arrangement found in nature, lipid bilayer tubes, remains difficult to study. 

Tubular membranes connect cells, facilitating intercellular transport and communication, e.g. axons and dendrites of neurons\cite{DebannePR11}. They also exist as protruding filaments involved in adhesion and cell movement,\cite{Lee2023} and as cilia to move fluids, propel cells, and form sensory receptors\,\cite{WoodTCB15}. Within the cell, tubes are found in organelles such as the Golgi apparatus \cite{RouxSM13} and the \ac{er} \cite{ChenCOCB13}. They form part of a dynamic network including flattened vesicles, called cisternae, in which tubes are remodelled over time, while others are static with nodes where multiple tubes intersect \cite{LinBJ14}. New tubes in these organelles are often formed by mechanical interactions of the membrane with the cytoskeleton\cite{RouxSM13}, and likewise tubes in model membrane systems tend to be formed from planar membranes in response to external perturbation. Such perturbations may be physical, such as by pulling\cite{EvansS96} or compressing\cite{StaykovaPNAS11} the membrane, or chemical, such as osmotic gradients over the membrane\cite{BhatiaSM20} or depleting lipids from one leaflet\cite{RahimiBJ16}.

Reported model systems for bilayer tubes start from \acp{GUV}, for example by adhering a patch of the \ac{GUV} outer membrane to a solid substrate by biotin-avidin binding and pulling the \ac{GUV} away \cite{EvansS96}, leaving a tubular connection between the \ac{GUV} and the patch bound to the substrate. Other works created two \acp{GUV} connected by a tube by splitting a \ac{GUV} using carbon fibres controlled by micromanipulators \cite{KarlssonN01,StepanyantsNL12}. While tubular membranes formed in these ways have a controlled length, their free-standing nature limits the range of characterisation techniques that can be applied. For example, surface-sensitive methods commonly used to study lipid bilayers, such as \ac{AFM}, are difficult to implement on unsupported structures. Supported tubular membranes therefore offer a promising alternative. 

Straight supported membrane tubes, tens of microns in length, have previously been produced by flow-induced extrusion from dry lipid films on passivated glass surfaces and used to study membrane fission\cite{DarNP17}. In our earlier work,\cite{ReganL19} we showed that spin-coating of a dry lipid film onto a glass substrate followed by gentle humidification can spontaneously generate extensive tubular networks. These networks remain surface-attached and, once formed, are stable over timescales of minutes to hours, making them well-suited for studying tubular bilayer structures using a wide range of experimental approaches. We demonstrated that \ac{qDIC} is a non-invasive optical microscopy technique suited to quantitatively image nanoscale structures, such as lipid bilayer thicknesses and single nanoparticles sizes, with sub-nm precision.\cite{ReganL19,HamiltonA22}

Building on these earlier studies, here we investigate the formation of supported tubular networks using controlled lipid types, namely \ac{DOPC} and \ac{dcpc}, exhibiting a liquid disordered (\Ld) and a solid ordered (\So) phase at room temperature, respectively. We apply \ac{qDIC} to quantify the tube radii over hundreds of tubes, for statistical significance. Beyond sizes, \ac{qDIC} is sensitive to birefringence, providing information on the three-dimensional spatial arrangement of tubular structures below the optical resolution, and various geometries are revealed. A biophysical mechanism is proposed for the formation of these structures, regulated by surface tension and lipid exchange with connected supported lipid bilayers. 

\section{Materials and Methods}

\subsection{Chemicals}

Lipids without fluorescent labels, namely \ac{DOPC} and \ac{dcpc}, were obtained from Avanti Polar Lipids (Alabaster, US) either as powder or pre-dissolved in chloroform, and used without further purification. For fluorescence, \ac{DOPE} labelled with ATTO488 was obtained from ATTO-TEC (Siegen, Germany). Chloroform, acetonitrile and 2-propanol were obtained from Sigma-Aldrich (St Louis, US) at HPLC grade or above.

\subsection{Spin Coating}\label{spincoating}

\Acp{slb} were prepared on glass coverslips using a spin-coating procedure based on that developed by \citet{MennickeL02}, and described in our previous work\cite{ReganL19}.

24$\times$24\,mm$^2$ \#1.5 coverslips (Menzel-Gl\"{a}ser, Germany) were cleaned by first wiping with acetone-soaked cleanroom paper, and then, to remove any remaining organic material and render the glass surface hydrophilic, etching in a 3:1 (v/v) solution of sulphuric acid to 30\% hydrogen peroxide (Piranha solution) at 95\textdegree C for one hour. Finally the coverslips were rinsed in distilled water, dried under nitrogen flow, and then stored under nitrogen at 4\,\textdegree C for no more than two weeks before use. 

Immediately before spin coating, lipid solution was added to the centre of the coverslip, which spread out to fully wet the glass surface. The type of solvent affects lipid bilayer formation by spin coating, and was chosen to optimise bilayer formation\cite{Regan19}.  For lipids forming a homogeneous \Ld\ phase bilayer at room temperature, a 95:5 (v/v) mix of chloroform and acetonitrile was used, while for lipid mixtures that formed either \So\ or liquid ordered (\Lo) phases at room temperature, 2-propanol was used. The solution volumes were chosen to fully wet the surface and depend on the hydrophilicity of both the solvent and glass substrate, being typically around 300\,\textmu l for  chloroform:acetonitrile mixture and 150\,\textmu l for 2-propanol. 

Spin coating was carried out in air on a Laurell WS-650-23 spin coater, using 30\,s at 3000\,rpm, with 6\,s acceleration and deceleration steps, leaving a dry lipid film on the coverslip. The lipid concentrations used were between 0.8 and 1.2\,mg/ml, adjusted for each mixture to produce lipid films that were mainly unilamellar after hydration, with some gaps and bilamellar regions. The lipids incorporated 0.1 mol\% ATTO488-\ac{DOPE} for fluorescence imaging unless otherwise stated. A detailed discussion of the development of these parameters is given in \Onlinecite{Regan19}. 

After spin coating, the lipid-coated coverslips were subjected to a pre-hydration step, in which they were incubated in a 100\% humidity nitrogen environment at 37\,\textdegree C for one hour. Pre-hydration was used to encourage lipid bilayer formation on the surface, without direct contact to bulk liquid water or flow, in order to avoid washing off lipids into liquid volume. Next, lipid-coated coverslips were allowed to cool to room temperature in a 100\% humidity nitrogen environment for approximately one minute. Finally, the lipids were fully hydrated by gently placing the lipid coated face of the coverslip onto a slide with an adhesive gasket filled with room temperature \ac{PBS} solution of 1$\times$ concentration, providing a liquid filled chamber of 120\,\textmu m height, in which reorganisation of the lipids completed.

\subsection{Imaging}

Samples were imaged as previously described.\cite{ReganL19} Briefly, a modified Nikon Ti-U inverted microscope with a 0.75\,NA 20$\times$ objective (Nikon CFI Plan Apo Lambda MRD00205) and 1.5$\times$ tube lens were used, providing a  30$\times$ magnification on the camera. Images were acquired using a Hamamatsu Orca 285 CCD camera (sensor size $1344 \times 1024$, pixel size 6.45\,\textmu m, full well capacity 18\,ke, read noise 7 electrons, 4.6 electrons per count at zero gain, 12-bit digitisation), yielding a corresponding pixel size on the sample of 216\,nm. The microscope stage was enclosed in a sealed incubation box connected to a Life Imaging Systems Cube2 heating unit, allowing the temperature at the sample to be controlled;  measurements were taken at room temperature unless otherwise stated.

For fluorescence measurements, a metal-halide high pressure lamp (Prior Lumen 200) was used with a epi-fluorescence illuminator (Nikon Ti-FL), filter turret (Nikon Ti-FLC), and a filter cube (Semrock GFP-A-BASIC, excitation 35\,nm wide centred at 469\,nm, emission 39\,nm wide centred at 525\,nm).  For \ac{DIC}, a 100 W halogen lamp (Nikon Ti-DH D-LH/LC) was used with a green interference filter (Nikon GIF) to provide a illumination wavelength centred at 550\,nm and a full-width at half maximum (FWHM) of 53\,nm. In addition, a Schott BG40 filter was used to remove wavelengths above 650\,nm which were not blocked sufficiently by the GIF. The illumination was then transmitted through a de-Sénarmont compensator (a rotatable linear polariser and quarter-wave plate, Nikon T-P2 DIC Polariser HT MEN51941) and focussed by a 0.72\,NA condenser (Nikon CLWD MEL56100) onto the sample. We used Nikon N2 prisms with a shear separation of $s=238\pm10$\,nm (a MBH76220 slider for the objective and a MEH52400 module in the condenser).  For \ac{qDIC} analysis, pairs of \ac{DIC} images were taken, with the angle of the linear polariser set to $\pm\theta$, with $\theta$ being either $13$\textdegree\ or $15$\textdegree. 

\subsection{Quantitative DIC}

\Acf{qDIC} allows the generation of quantitative phase maps from pairs of standard \ac{DIC} images, as detailed in our previous works \cite{McPheeBJ13,ReganL19,HamiltonA22}. Briefly, a pair of \ac{DIC} images is taken at opposite de-Sénarmont polariser angles ($I_{+}$ at $+\theta$ and $I_{-}$ at $-\theta$), which are then used to produce a contrast image
\begin{equation}
\Ic = \frac{I_{+} - I_{-}}{I_{+} + I_{-}}\,,
\label{eq:DICcontrast}
\end{equation}
from which the phase difference in the sample $\delta$ can be retrieved using 
\begin{equation}
\Ic = \frac{\sin(2\theta)\sin(\delta)}{\cos(2\theta)\cos(\delta) - 1}\,,
\label{eq:DICcontrast2}
\end{equation} 
which can then be rearranged\cite{HamiltonA22} into an analytical expression to determine $\delta$ from $I_{\rm{C}}$ and $\theta$. The quantitative phase map $\phi(\br)$ of the sample is then found by integrating $\delta(\br)$ along the shear direction with a Wiener filter using a signal-to-noise parameter $\kappa$ which is creating a high-pass cut-off in spatial frequency\cite{HamiltonA22}. To reduce the striping resulting from this integration, a minimisation algorithm penalising spatial gradients can be used, as shown in \citet{ReganL19} and Section\,\ref{S-sec:gradmin}.

\section{Results and Discussion}

\subsection{Bilayer structures of \ac{DOPC} lipids}

\begin{figure*}
	\includegraphics[width=\textwidth]{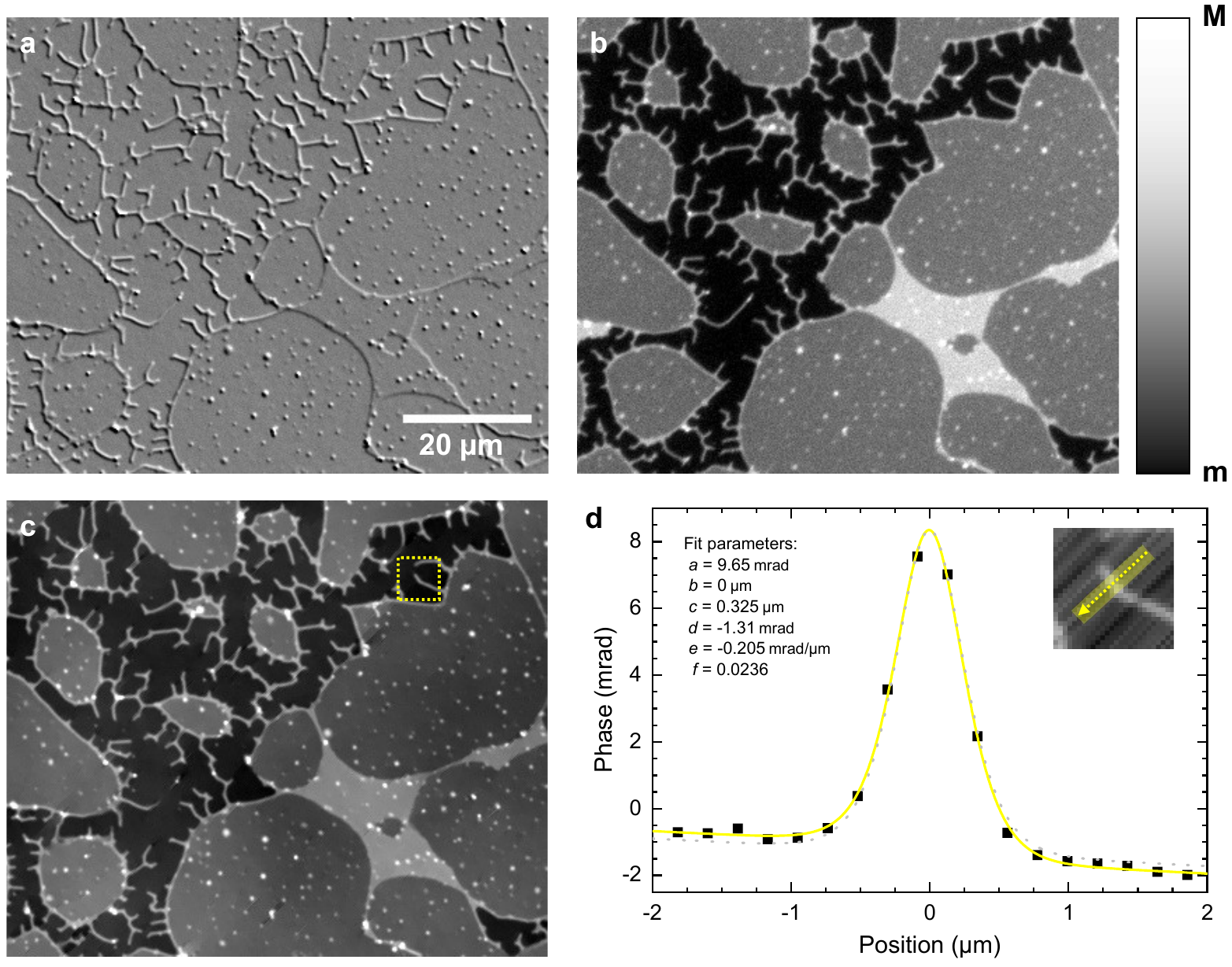}
	\caption{a) A \ac{qDIC} $\delta(\br)$ image showing a region of a spin-coated \ac{DOPC} \ac{slb} stack in which tubular networks have formed, scaled from $m$ = $-6$\,mrad to $M$ = $+6$\,mrad. b) The corresponding fluorescence image, scaled from $m$ = 80 to $M$ = 260 counts.  c) The same region shown as a minimised \ac{qDIC} phase $\phi(\br)$ image, scaled from $m$ = -12\,mrad to $M$ = 8\,mrad. d) The phase profile across a tube; the line cut is shown in the inset. The yellow line is a fit using \Eq{eq:birefringentfit} with the parameters shown, and the grey dashed line shows the fit function using $f=0$, i.e. without birefringence.}
	\label{fig:DOPCBilayerTubes}
\end{figure*}

As introduced in our earlier work,\cite{ReganL19} spin-coated lipid films form thin ``branch-like'' lipid networks on the glass surface upon hydration, in addition to planar lipid bilayers. These networks form spontaneously from spin-coated lipid films, and an example in a supported \ac{DOPC} lipid bilayer stack is given in \Fig{fig:DOPCBilayerTubes}, showing the phase difference $\delta(\br)$ in (a), reconstructed phase $\phi(\br)$  in (c), and fluorescence in (b). A detailed discussion of \ac{qDIC} images characterising the planar bilayers was given in our earlier work.\cite{ReganL19} Here, we focus on the characterisation of the tubular structures. From \Fig{fig:DOPCBilayerTubes}, we observe that these structures are connected
to unilamellar planar bilayer regions (see mid-range grey contrast), and exhibit a bright contrast, similar to that of a bilamellar stack.

To extract quantitative structural information, we take line profiles perpendicular to the tubular's direction,\cite{ReganL19} as illustrated in \Fig{fig:DOPCBilayerTubes}c-d. To reduce the noise in these cross-section profiles, we average over a width of four to eight pixels ($0.9$ to $1.7$\,\textmu m). In order to minimise the influence of \ac{qDIC} integration artefacts, we only take cross sections that are nearly parallel to the \ac{DIC} shear direction (further discussion on integration artefacts is provided in our earlier works\cite{ReganL19,HamiltonA22}). These cross-sections are then fitted using 
\begin{equation}
t(x)=a\left[\sech^2\left(\frac{x-b}{c}\right) + f\tanh\left(\frac{x-b}{c}\right) \right] + d + ex\,,
\label{eq:birefringentfit}
\end{equation}
where the $\sech^2$ term represents a structure which is smaller than the optical resolution of our imaging system, thus appearing as a peak in the phase profile. The parameter $a$ is the peak amplitude, $b$ is the peak position, and $c$ the peak width giving a measure of the tube diameter convoluted with the optical resolution. Local background gradients in the image are taken into account by the linear term $d + ex$. The $\tanh$ term with relative amplitude parameter $f$ models the effect of birefringence, as detailed later. 

An example of a phase cross-section through a tubular structure, and the corresponding fit, is shown in \Fig{fig:DOPCBilayerTubes}d. By integrating the area contained within the curve, we can quantify the amount of bilayer material in the cross-section. The integral of the $\sech^2$ term is $2ac$. We can convert this value into a corresponding bilayer length, dividing by the optical thickness of a single bilayer, which is 5.73 mrad as determined in our previous work \cite{ReganL19}. For the structure shown in \Fig{fig:DOPCBilayerTubes}d, we find a lipid bilayer length of $(1096 \pm 57)$\,nm in cross-section, with the uncertainty derived from the 95\% confidence interval of the fit. Attributing the observed structure to a tube geometry, yielding a radius of $(174 \pm 9)$\,nm, is consistent with its resolution limited appearance. The alternative configuration consisting of a stack of two flat bilayers would require a width exceeding 500\,nm to contain this amount of lipid, while the peak width in the fit in \Fig{fig:DOPCBilayerTubes}d is resolution limited ($c=325$\,nm). 

Further evidence for the tubular geometry of the membrane is provided by the slight mismatch of the background level on either side of the peak in the phase profile. This is modelled by the $\tanh$ term in \Eq{eq:birefringentfit}, with the prefactor $f$ describing the relative amplitude of the step. The scale of this effect is illustrated by the grey dashed line showing the fit with $f$ set to zero.
Such changes in phase around the peaks can be explained by the birefringence of the lipid bilayers, which have an ordinary refractive index (\nor=1.445) for light polarised in the bilayer plane, and an extraordinary index (\nex=1.460) for light polarised perpendicular to that plane.\cite{DevanathanFJ06}. In \ac{DIC}, the two sheared beams are polarised along and orthogonal to the shear. In a tubular geometry, the bilayer is oriented partly out of plane, so that for tubes aligned orthogonal to the shear, one of the beams is polarised along the tube, and thus experiences only \nor, while the other is polarised orthogonal to the tube, and thus experiences a mixture between \nor\ and \nex, depending on tubular geometry. This results in a phase difference between the two beams, which, after integration, provides a phase step. Notably, this effect does not occur for planar bilayer geometries, where both \ac{DIC} polarisations are in-plane. Hence, the observed phase-step provides evidence for the tubular geometry. 

\begin{figure*}
	\includegraphics[width=\textwidth]{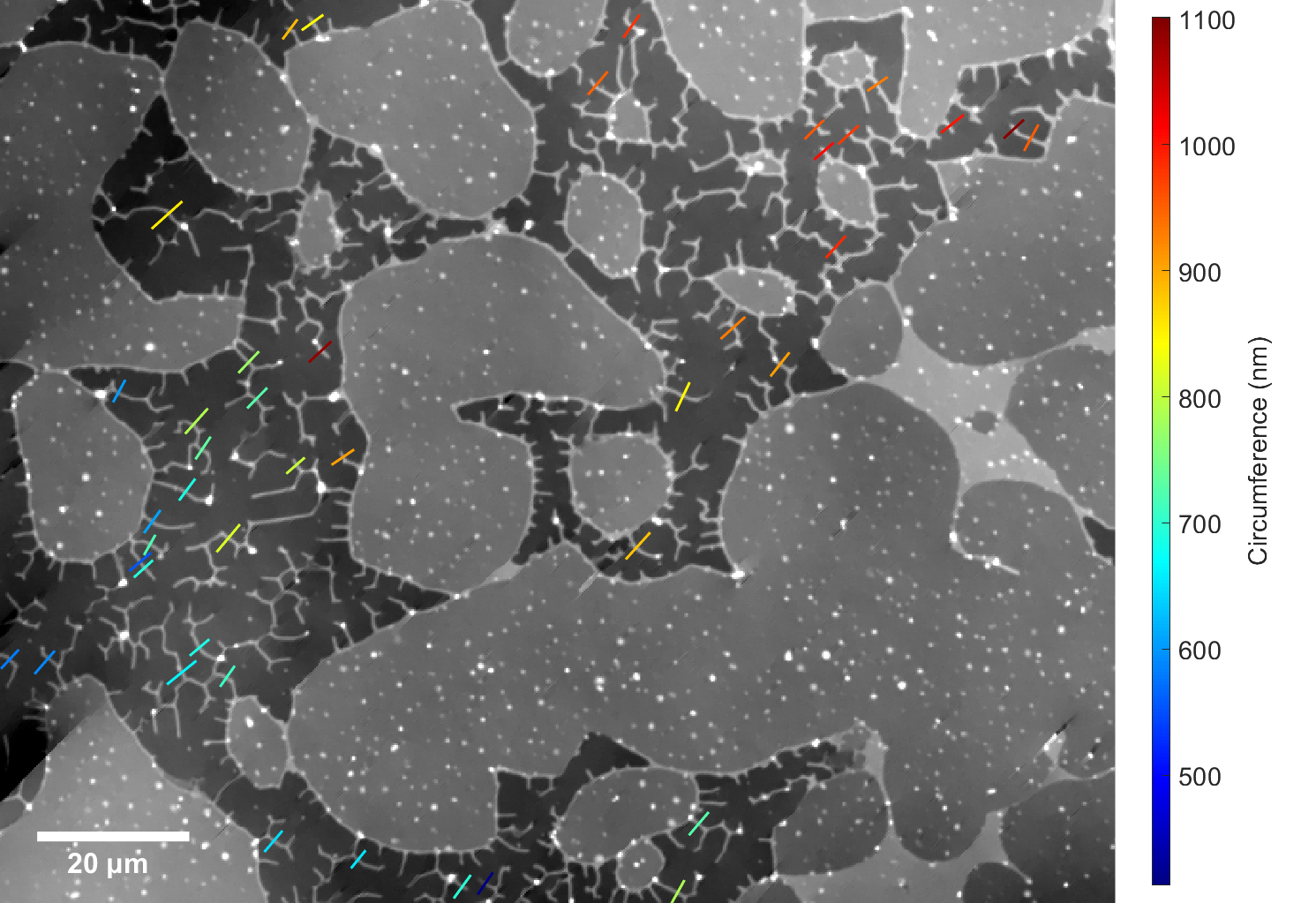}
	\caption{Minimised \ac{qDIC} phase image containing the region shown in \Fig{fig:DOPCBilayerTubes}, scaled from -18 mrad to 6 mrad. Lines show measured profiles across tubes, with a colour encoding the resulting tube circumference on a scale as indicated.}
	\label{fig:TensionVariationEM}
\end{figure*}

\begin{figure*}
	\includegraphics[width=\textwidth]{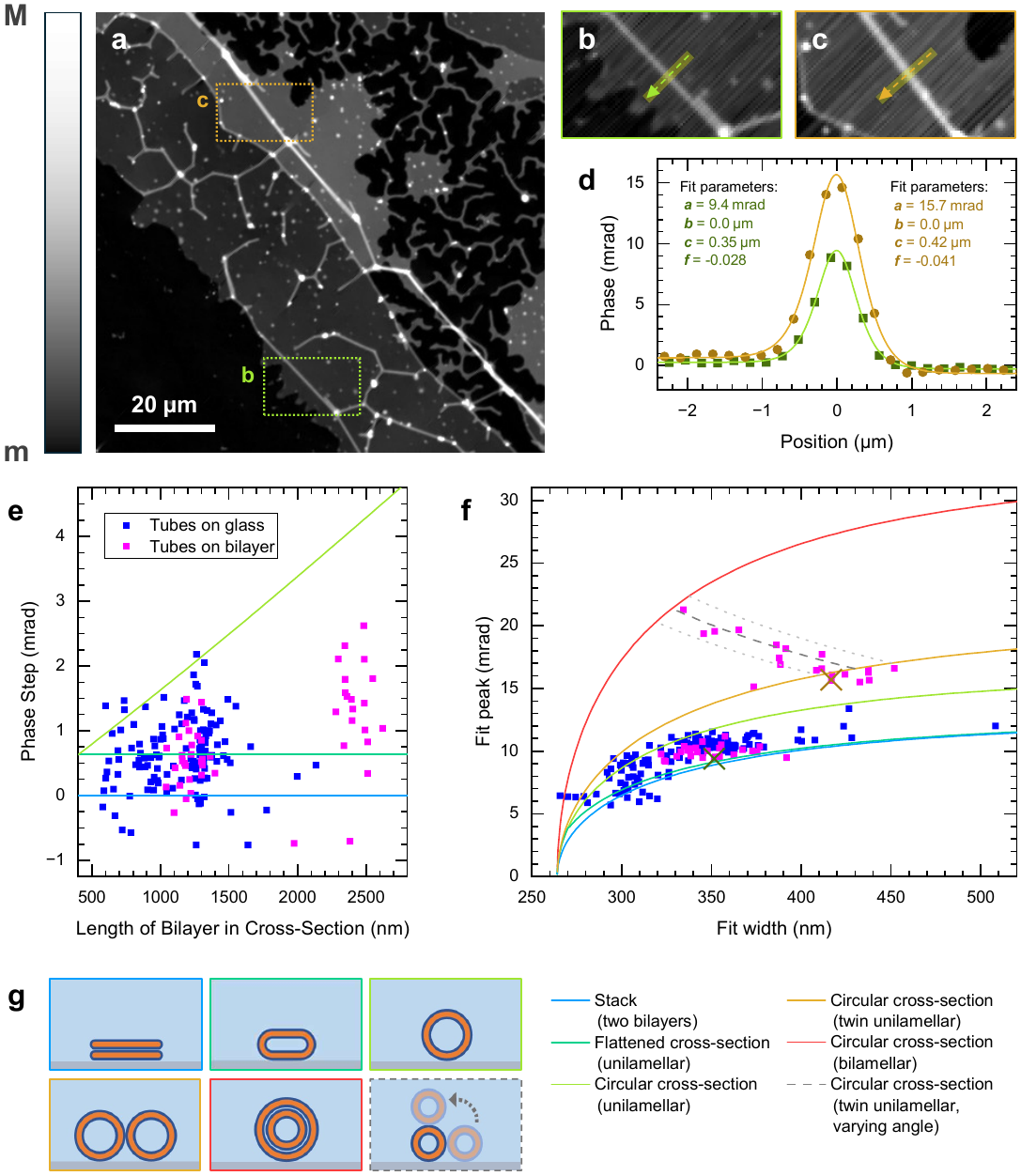}
	\caption{a) Minimised \ac{qDIC} phase image showing a region with tubes formed on top of other bilayers, as well as tubes formed on glass, scaled from $m$ = 0\,mrad to $M$ = 32\,mrad. b,c) Line cuts through two different tubes, with phase profiles and fits shown in d). e) Phase step $2fa$ due to bilayer birefringence versus the length of the bilayer cross section for tubes formed on glass (blue) and on other bilayers (magenta). Coloured lines show the expected phase step for different  cross-sectional shapes (see upper row of g) as a function of circumference. f) Fitted peak amplitude $a$ against width $c$, from the fitting to the \ac{DOPC} tube phase profiles, with simulated curves for different cross-section shapes shown as coloured lines. The positions of the points corresponding to the phase profiles shown in d) are denoted by crosses. g) Illustration of the different cross-section shapes simulated in e) and f), with corresponding frame colour. }
	\label{fig:BilamellarTubes}
\end{figure*}

Over a larger region shown in \Fig{fig:TensionVariationEM}, we observe tubular structures of different contrast, indicating a variation in tube size. We fitted tube cross-sections as in \Fig{fig:DOPCBilayerTubes}d over this field of view, along the coloured lines, and similarly over more fields of view (see \Sec{S-sec:MeasurementPositions}), and provide in \Fig{fig:BilamellarTubes}e the resulting phase steps as function of tube circumference and in \Fig{fig:BilamellarTubes}f the peak amplitude as function of the width. Assuming that the tubes are unilamellar with circular cross-sections, the mean and standard deviation of the radii is $181.8\pm47$\,nm, over $n=125$ measurements. The mean diameter of 364\,nm is close to the diffraction limit, consistent with our fit. The radii of these supported tubes are similar to those of tubes pulled from \acp{GUV}\cite{KarlssonN01}, but are larger than the typical radii of tubules in cellular organelles, which tend to be below 100\,nm\cite{RouxSM13}. To understand these differences, the following considerations can be made. 

For the case of a tube with a circular cross-section, the radius $r$ is determined by the balance between the bending rigidity of the bilayer, $\eta$, and the tension in the tube, $\sigma$, resulting in\cite{RouxSM13} 
\begin{equation}
r = \sqrt{\frac{\eta}{2\sigma}}\,.
\label{eq:RadiusEq}
\end{equation}
The bending rigidity depends on the lipid species present, and on the relative distribution of these lipids between the two leaflets of the bilayer. Our system is almost exclusively formed from phosphocholine (PC) lipids which have a low spontaneous curvature\cite{KurczyPLOS114}. In contrast, biological membranes often contain curvature inducing proteins\cite{RouxSM13}, as well as significant proportions of lipids with high intrinsic curvature such as phosphoethanolamines (PE)\cite{EscribaJCMM08,SprongNRMCB01}, which provide finite equilibrium radii $r_0$, yielding
\begin{equation}
	r^{-1} = \sqrt{\frac{2\sigma}{\eta}}+r_0^{-1}\,.
	\label{eq:RadiusEq2}
\end{equation}
Indeed, it has been previously shown for tubes pulled from \acp{GUV} that when tubes can draw from reservoirs of higher spontaneous curvature PE lipids, tube diameters are reduced\cite{KurczyPLOS114}.  

The surface tension in the tubes is governed by the overall sample geometry and the membrane-substrate interactions. Comparable model systems to ours consisting of linear supported membrane tubes produced by flow-induced extrusion have radii in the $10$ to $40$\,nm range\cite{DarNP17}, considerably smaller than what we observe with similar lipid compositions, suggesting that our system is under lower tension. With \ac{qDIC} we can explore tension variations within our system, since for a constant lipid composition (and therefore constant $\eta$), variations in tube radius must correspond to changes in tension. In \Fig{fig:BilamellarTubes}e, we clearly see different groupings of circumference values in the data, with clustering around 650\,nm, 950\,nm, and 1300\,nm, which correspond to measurements taken in different sample regions.Even within single fields of view, such as shown in \Fig{fig:TensionVariationEM}, we can see clustering of different circumference values; in the bottom left corner we see tubes with lower circumference values corresponding to the 650\,nm cluster, while the tubes in the upper right belong to the 950\,nm cluster. This indicates that the tube circumference reports variations in membrane tension over ranges of just a few hundred microns within the \ac{slb} (see also supplementary material, \Fig{S-fig:TensionVariation}).

In addition to the short, branched tubes protruding from the edges of the bilayer stack, as seen in \Fig{fig:DOPCBilayerTubes}, we observe tubes with long straight sections. An example is shown in \Fig{fig:BilamellarTubes}a where a structurally distinct tubular network sits on top of a \ac{slb} band extending from top left to bottom right. These tubes appear to be above rather than below the bilayer, as the latter would result in the bilayer bending around them, which could bee seen as a increased optical thickness. Those tubes have straight segments often over 10\,\textmu m in length, compared to a few \textmu m lengths observed for tubes on glass, suggesting that the tube-substrate interaction has a major role in shaping the tube networks. We attribute the shorter tube lengths on glass to adhesion points on the glass surface which create corners or tips in the tubular network. These are are screened in bilayer-coated regions, allowing for long straight tubes to be created by the surface tension.

The results of fits to the tubes formed on top of other bilayers are shown in \Fig{fig:BilamellarTubes}e as magenta points, and we find also here clustering of the data around specific circumference values. The lower cluster of points has a mean circumference of $1213$\,nm and a standard deviation of $90$\,nm ($n = 27$), which is similar to the circumference of the tubes on glass in the same region ($1332$\,nm, standard deviation $161$\,nm, $n = 66$). This finding suggests that both types of tubes are under similar levels of tension, as expected if they are formed by a contiguous bilayer.

The population of higher circumference tubes centred around 2400\,nm would have a diameter of about 800\,nm, assuming a unilamellar tube with a circular cross-section, which should be resolvable in our qDIC microscope. An example of one of these tubes is given in \Fig{fig:BilamellarTubes}c, with the corresponding phase profile and fit shown in orange in \Fig{fig:BilamellarTubes}d. For comparison, one of the lower circumference tubes is shown in green (\Fig{fig:BilamellarTubes}b). While there is an increase in the fit width, $c$, of the tube in \Fig{fig:BilamellarTubes}c (417\,nm, compared to 352\,nm for the tube in \Fig{fig:BilamellarTubes}b), it is less than would be expected for a unilamellar tube of 800\,nm diameter. 

We must therefore consider alternative cross-section tube configurations, illustrated in \Fig{fig:BilamellarTubes}g. We note that the fit parameters of the \ac{qDIC} phase profiles, specifically the peak amplitude $a$, width $c$, and the phase step $2fa$, are dependent on the tube geometry. To identify different tube geometries, we therefore calculated the expected phase step due to the birefringence for different cross-sections, shown as coloured lines in \Fig{fig:BilamellarTubes}e. We also simulated spatial phase profiles neglecting birefringence, and applied \Eq{eq:birefringentfit}. Details are given in \Sec{S-sec:tubephasemodel}. The resulting peak amplitudes and widths are shown in \Fig{fig:BilamellarTubes}f, for a non-tubular arrangement of two stacked bilayers (blue line), a flattened unilamellar tube cross-section using an edge radius of 64.8\,nm (dark green line), a unilamellar tube (light green line), a pair of unilamellar tubes running side-by-side (orange line), and a bilamellar tube with a circular cross-section (red line).  We can see that the majority of the data for the tubes formed on glass reside between the simulation of a unilamellar flattened bilayer cross-section and that of a circular cross-section, suggesting a slightly squashed configuration of the membrane. We note that a slight defocus of the images can lead to an increased width, which might explain the data points below the bilayer stack simulations. Flattened membranes show little differences to two stacked bilayers when the cross-section is dominated by the flat part. Yet, these two configurations correspond to distinct predictions regarding the phase step due to the birefringence (\Fig{fig:BilamellarTubes}e), indicating that with \ac{qDIC} we can distinguish these cases. Notably the birefringence data resides between that of a circular crosssection (green line) and a flattened geometry (dark green line). 

\begin{figure}
	\includegraphics[width=\columnwidth]{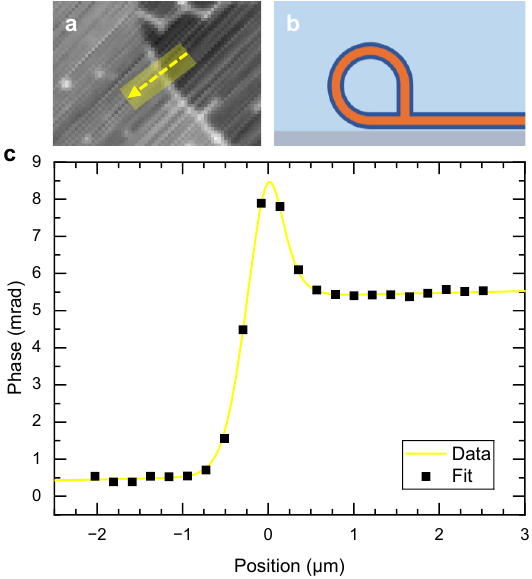}
	\caption{a) Phase image of a unilamellar region with a border of higher optical thickness. b) Illustration of the proposed cross-section topology at the border, with a bilayer triple junction structure. c) Phase profile along the region indicated in a), and fit.}
	\label{fig:EdgeStructure}
\end{figure}

In \Fig{fig:BilamellarTubes}f we see that the tubes formed on top of other bilayers (magenta points) are split into two populations. One population is comparable to the tubes formed on glass in terms of shape, suggesting that these types of tubes likely share a common structure. The other population has significantly higher peaks than expected for a unilamellar bilayer with circular cross-section, but not as high as would be expected for a bilamellar circular cross-section where the inner tube is in direct contact with the outer tube. Notably, these data points appear to follow a trend in \Fig{fig:BilamellarTubes}f. The orange line shows the case of two circular tubes of equal radius attached to each other flat on the surface (illustrated in \Fig{fig:BilamellarTubes}g), and appears to act as lower bound for these data points. The dark grey dashed line represents a pair of attached circular cross-section tubes with a radius given by the mean of the isolated tubes (193\,nm), with varying orientation from side-by-side to on top of each other. The lighter grey dashed lines are simulated with radii at plus and minus one standard deviation from the mean radius (207\,nm and 179\,nm). Thus, \ac{qDIC} indicates that this population of tubes is not bilamellar, but consists of pairs of tubes wrapped around each other, like a twisted pair cable (see \Fig{S-fig:TubePosition4} for measurement positions  ).

Beyond characterising the tubes themselves, it is interesting to investigate the lipid structure at the point where tubes meet the planar supported bilayer regions of the sample. We see in \Fig{fig:DOPCBilayerTubes} that the unilamellar bilayer regions from which the tubular networks protrude appear to be surrounded by thicker borders running along their perimeter, visible in fluorescence (\Fig{fig:DOPCBilayerTubes}b) and qDIC phase (\Fig{fig:DOPCBilayerTubes}c) as lines of higher contrast. We can fit the phase profile over this border using \Eq{eq:birefringentfit}. In this context, the $\tanh$ term is dominated by the step from the optical thickness of the planar lipid bilayer regions, rather than the birefringence. Thus, substituting the position $b$ in the $\tanh$ term with an additional parameter $g$, allows us to fit the positions of the step-edge of the bilayer patches independently of the position of the border tubular structure.

An example of a phase profile over a unilamellar edge border is shown in \Fig{fig:EdgeStructure}. We see a clear step change in the background due to the transition from the bare glass to a single planar bilayer, and a sharp peak at the edge. We find in these fits that the mean difference between the position of the border and the bilayer edge is zero within error, suggesting that the two are physically linked in some way. Furthermore, we find that the mean length of extra bilayer lipid contained within these borders is $(666\pm55)$\,nm ($n = 24$, error given is the standard error), which is smaller than that of the tubes they connect to in the same region of interest ($(861\pm21)$\,nm, $n = 49$). We propose that the edge structures are tubular, involving a bilayer triple junction, as illustrated in \Fig{fig:EdgeStructure}b.
In this geometry the tubular structure misses about one quadrant of a complete tube circumference, which would be consistent with the observed $(77 \pm 7)$\% relative bilayer length calculated from the above given lengths. It would be interesting to use simulations to get a prediction of the structure including the relative length to compare with the experimental results.

\subsection{Bilayer structures of \ac{dcpc} lipids across the phase transition}\label{sec:TubeFormation}

\begin{figure*}
	\includegraphics[width=0.9\textwidth]{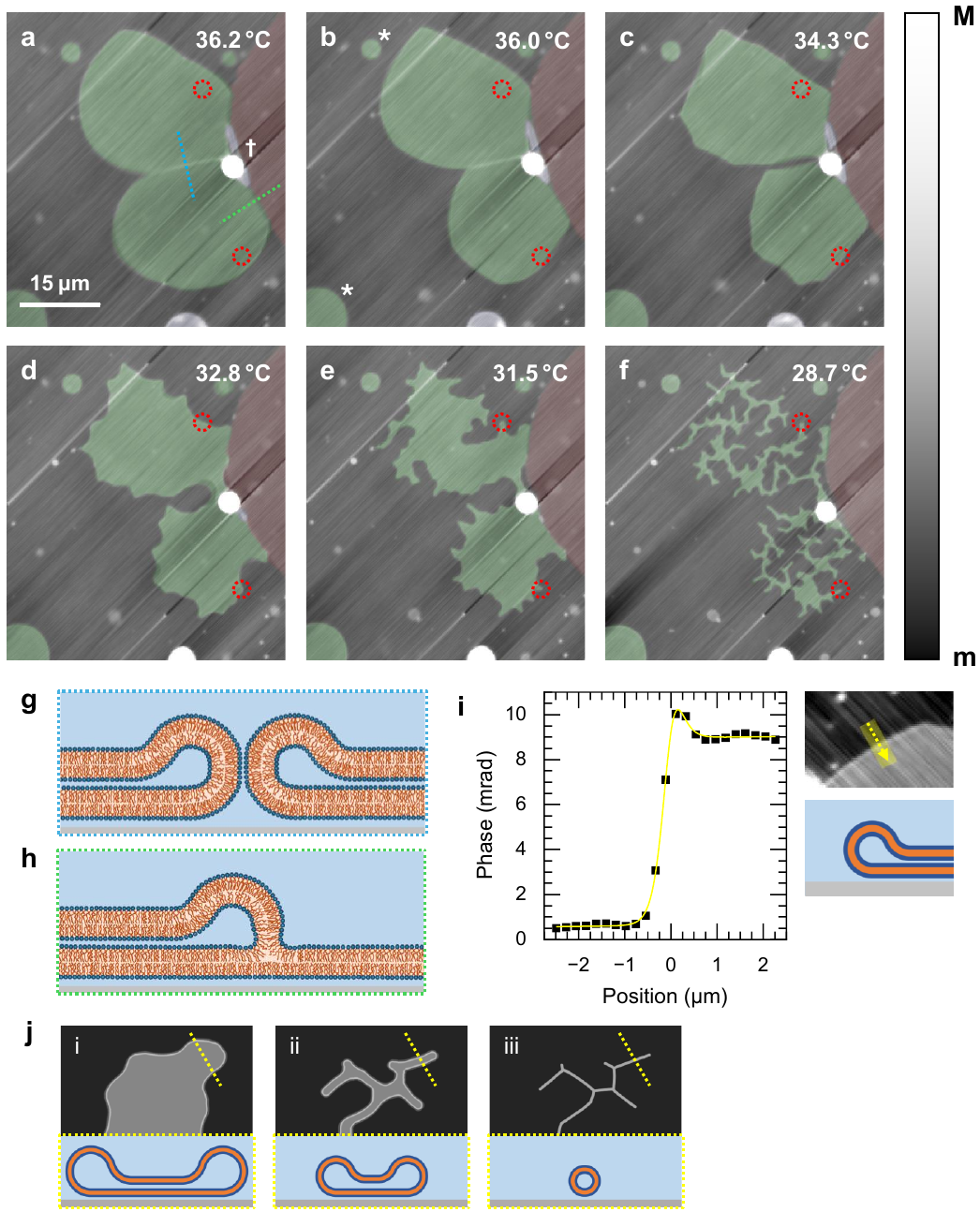}
	\caption{a–f) \ac{qDIC} phase images showing shape changes in a \ac{dcpc} bilayer during cooling ($m = -33$\,mrad, $M = 23$\,mrad). False colour indicates lamellarity: unilamellar (red), bilamellar (green), and multilamellar (blue). Red circles mark selected membrane–surface pinning points. g,h) Illustrations of bilayer arrangements at locations marked by blue and green dashed lines in (a). i) Phase profile across a bilamellar edge with structural illustration.
j) Illustration of the mechanism of bilamellar-to-tubular rearrangement, shown as plan-view and cross-section.}
\label{fig:TubeFormation}
\end{figure*}

\ac{dcpc} bilayers have a phase transition from \Ld\ to \So\ at a temperature\cite{KoynovaBBA98} of $T_m = 33.7$\,\textdegree C, and are therefore in the \So\ phase at room temperature. The \So\ phase has a higher thickness and a higher areal density of lipid molecules than the \Ld\ phase. SLBs using \ac{dcpc} were prepared as explained in Materials and Methods, by prehydration at 37\,\textdegree, followed by cool-down over a few minutes and final hydration at room temperature. These samples were then mounted and observed in the qDIC microscope which was pre-heated at 37\,\textdegree C, hence above  $T_m$. After mounting, samples were allowed to stabilise for at least 10 minutes, and then the temperature was lowered, typically in 0.3\,\textdegree C steps with 10\,min intervals. An example of the resulting shape transformation occurring in the membrane bilayers, including the formation of tubular structures, is shown in \Fig{fig:TubeFormation}a-f.

In \Fig{fig:TubeFormation}a, a section of the bilayer stack is observed after it has been heated in the microscope above $T_{\rm{m}}$ to $36.2$\,\textdegree, as measured at the sample. Using the qDIC phase information, we identify a large bilamellar region (highlighted in green) attached to a unilamellar region on the right (highlighted in red). The bilamellar region is comprised of two separate parts, as indicated by the line of higher phase contrast at the intersection between the two. The existence of this boundary suggests that at the edges of these bilamellar parts, the bilayers are folding back onto themselves, as illustrated in \Fig{fig:TubeFormation}g. This folding prevents the two patches from merging, and can be quantified in the phase profile over the bilamellar patch edge. \Fig{fig:TubeFormation}i shows a clear peak in the phase, where the bilayer is oriented vertically, which then levels off, as expected for a stack of two bilayers. The bilamellar regions can therefore be considered as deflated vesicles sat on the surface, analogous to the cisternae in the \ac{er}. By integrating the fit in \Fig{fig:TubeFormation}i, we find that the peak contains 558 nm total lipid length; assuming a semi-circular cross-section, this would correspond to a radius of curvature of 178 nm.

\begin{figure*}
	\includegraphics[width=\textwidth]{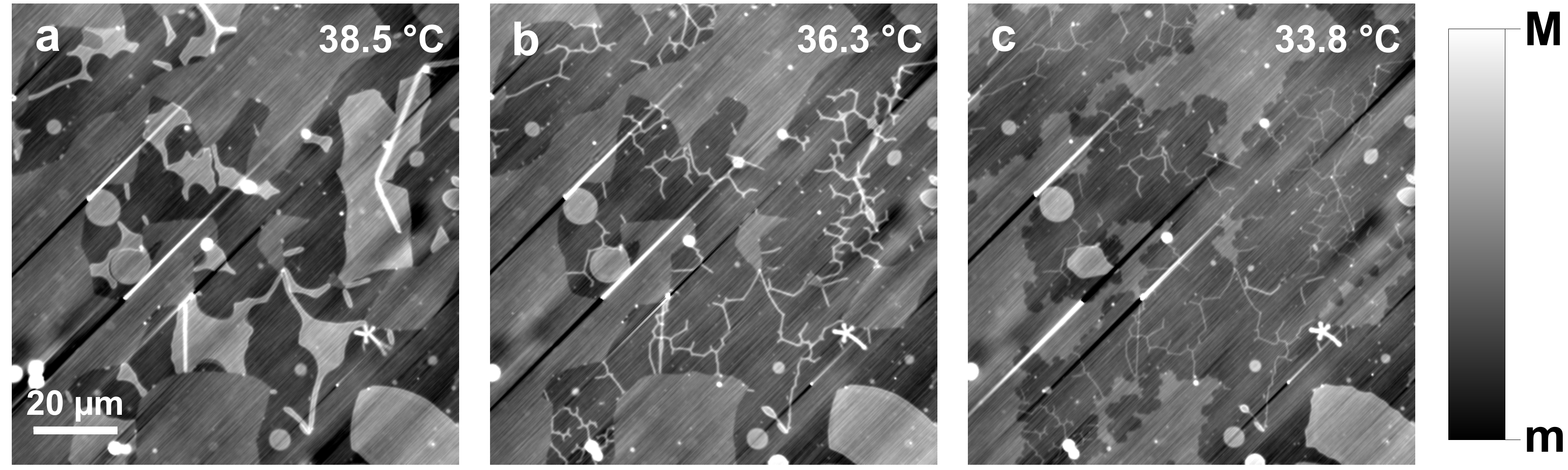}
	\caption{a–c) \ac{qDIC} phase images showing tube formation in an \ac{dcpc} bilayer sample during cooling ($m = -3$\,mrad, $M = 20$\,mrad). Note the different tube radii visible in c), suggesting different levels of tension in different tubular networks.}
\label{fig:TubeFormation3M}
\end{figure*}

As the sample is cooled (\Fig{fig:TubeFormation}b-f), both bilamellar regions gradually contract. Initially, the edges recede uniformly as the area of each bilamellar patch shrinks, but eventually the edges become `pinned' at certain points, giving them first an polygonal, then a jagged appearance. Two examples of such pinning points are indicated as coloured circles on \Fig{fig:TubeFormation}a-f. As the area of the bilamellar regions contracts further, more of these pinning points are encountered by the receding edge, until finally the bilamellar regions have completely reorganised into networks of flattened tubes linking these pinning points together. This process is illustrated in \Fig{fig:TubeFormation}j, which shows the transition from a bilamellar patch (i) to tubes (iii). A further example of this tube forming process is shown in \Fig{fig:TubeFormation3M}.

We also note that while the bilamellar regions shrink, the position of the boundary between the unilamellar and bilamellar regions remain mostly unchanged over time. In other words, the bilamellar regions do not themselves become unilamellar, but rather the two bilayers in the bilamellar regions contract together from the outer edge inwards. We therefore suggest that the structure at the unilamellar-bilamellar boundaries (see the green dashed line in \Fig{fig:TubeFormation}a) is a triple junction structure, as illustrated in \Fig{fig:TubeFormation}h, reminiscent of those we showed previously in \Fig{fig:EdgeStructure}. However, the observation that its position seems fixed is interesting, as one could expect that such a triple junction could move to reduce the douyble bilayer area. Relevant considerations are the enclosed water volume in the double bilayer, and the local geometry. Simulations of this situation would be helpful but seem not to be available.\cite{NoguchiSM12,AkimovMS20}

The observed tube formation mechanism resembles a spontaneous, two-dimensional analogue of methods which produce tubes from \acp{GUV} by attaching part of the \ac{GUV} surface to a solid substrate, and then physically pulling it away, leaving a membrane tube connecting the \ac{GUV} to the attachment point. This attachment of \ac{GUV} lipids to the substrate may be achieved for example by biotin-avidin binding\cite{EvansS96}. Interestingly, in our system, no additional molecules have been introduced to bind the membrane lipids to the glass substrate at particular sites, nor have we applied any surface passivation to prevent the tubes collapsing into supported bilayers\cite{DarNP17} or otherwise limit the tube-surface interaction. The effect observed is therefore driven purely by local variations in the interaction between the membrane and the hydrophilic glass surface.

Given that the force driving the tube formation shown in \Fig{fig:TubeFormation} is the depletion of lipid area from the bilamellar regions of the membrane, in the following we discuss what drives this depletion. Firstly, we note that bilamellar regions which are not connected to unilamellar regions generally do not undergo significant shape transformations during cooling. This is exemplified by the isolated bilamellar patches indicated by the asterisks in \Fig{fig:TubeFormation}b, which maintain a constant area. The amount of material in these patches can be quantified (see \Sec{S-sec:PhaseArea}). After subtracting a quadratic fit to the phase profile, accounting for background around each patch, and then integrating, we find the phase area of the patches in rad\,\textmu m$^2$. For the top-left and bottom-left patches, we find phase areas of $90.9 \pm 6.0$\,mrad\,\textmu m$^2$ and $893.0 \pm 10.3$\,mrad\,\textmu m$^2$ at 36.2\,\textdegree C, and $101.2 \pm 5.8\,$mrad\,\textmu m$^2$ and $974.1 \pm 7.8$\,mrad\,\textmu m$^2$ at 29.2\,\textdegree C, respectively.

Together, these observations indicate that bilamellar regions need to be connected to unilamellar regions for lipid transfer to occur, and suggest lipid migration from bilamellar to the unilamellar regions upon cooling. As a driving mechanism to this migration, we showed in our previous work that strong bilayer–surface interactions lead to stretching of the upper leaflet of the first bilayer \cite{ReganL19}. This stretching generates a tension, increasing with cooling due to the higher areal density of the \So\ phase requiring more lipids to cover the upper leaflet of the surface attached bilayer. Such tension is relieved by lipid migration from the upper leaflet of the upper lamella of the bilamellar regions flowing to the upper leaflet of the unilamellar region in the triple junction. In turn, bilamellar regions get gradually pulled off the surface from the edges, unless they are strongly anchored, for example in small cavities of the etched glass. These positions form the end or corner points of the remaining tubular structures.  

Interestingly, the constant areas of the unilamellar patches suggests that while lipid transfer into these regions occurs to compensate the increasing surface density of lipids during cooling in the upper leaflet, the surface attached are of the lower leaflet stays constant. This indicates that the surface interactions established during spin coating are significantly stronger than interactions after hydration, which is plausible due to the screening of electrostatic interactions by water. The area of the unilamellar regions are thus determined by the initial spin-coated configuration of the lipid film, and upon formation of bilamellar structure the lower leaflet can be stripped off the surface, as seen during the tube formation.

Another conceivable mechanism for the lipid loss in the bilamellar regions is a transfer into vesicles. In the \ac{dcpc} samples, we see small vesicles with sizes below the optical resolution, as well as some larger multilamellar vesicles. The latter can be close to the bilamellar regions undergoing rearrangement into tubes, such as seen in \Fig{fig:TubeFormation}a-f. However, these larger vesicles before and after the rearrangement of the bilamellar regions into tubes show no increase of their lipid content (see \Fig{S-fig:VesicleProfile}), ruling out significant transfer. 

The membrane phase transition influences the radius of a preformed tubular structure, which we have measured by monitoring the changes in the phase profile of tubes formed from a \ac{dcpc} bilayer stack as the sample is cooled below its main transition temperature, $T_{\rm m}$. For this study, we used a sample with a high surface coverage (approximately 72\% unilamellar, 20\% bilamellar, 2\% multilamellar) and a consistent overall structure. This uniformity helps minimise localised tension variations in the sample resulting from differences in membrane geometry, so any observed changes in phase profile can be attributed to global effects of the phase transition acting across the whole sample.

Importantly, the phase behaviour of this system is determined by the properties of a single lipid species, \ac{dcpc} (the effect of the 0.1 mol\% fluorescently labelled \ac{DOPE} is negligible). As such, tubes are expected to be entirely in the \So\ phase after cooling. However, previous studies on \acp{slb} on curved surfaces show that curvature on the order of tens to hundreds of nanometres can lower $T_{\rm m}$ by several degrees \cite{AhmedL09,BrummBJ96}, hence the phase transition in the tubes may occur at a few degrees lower temperatures.

\begin{figure}[t]
	\includegraphics[width=\columnwidth]{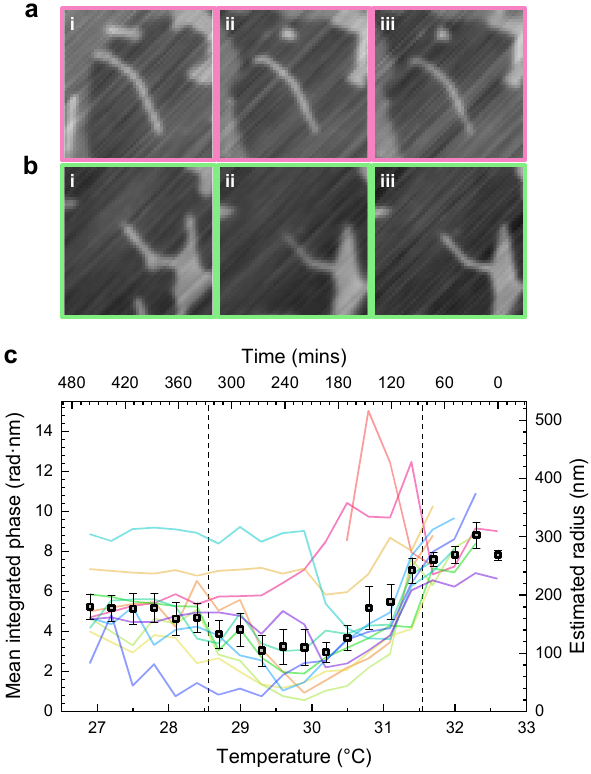}
	\caption{a,b) Examples of two tubes at three different temperature points, i) 31.7\,\textdegree C, ii) 29.3\,\textdegree C, iii) 27.5\,\textdegree C. qDIC phase from -14 to 14 mrad, image size 10.8\,\textmu m square. c) Integrated phase of twelve \ac{dcpc} tubes against temperature as the sample is cooled below its main phase transition temperature (the phase coexistence region in the \ac{slb} is indicated by the dashed lines) The grey squares show the average for tubes remaining connected to the bilayer stack. The right axis gives the tube radius, assuming \Ld\ phase parameters. The temperatures shown are the target temperature of the heating chamber. Error bars are standard deviations over the tubes.}
\label{fig:PhaseAreaVsTemp}
\end{figure}

\Fig{fig:PhaseAreaVsTemp} shows the integrated qDIC phase of twelve different tubes during the phase transition as the sample is cooled in 0.3\,\textdegree C steps every 24 minutes. The temperature range over which the phase transition of planar bilayers occurs is visible in the corresponding fluorescence images (see \Fig{S-fig:TubeRadii}) by the selective partitioning of the fluorophore into the \Ld\ phase coexisting with the \So\ phases during the transition in the planar regions of the bilayer stack, and is delimited in \Fig{fig:PhaseAreaVsTemp} by dashed lines. The integrated qDIC phase quantifies the evolution of the tube structure as the bilayer membrane changes from the \Ld\ to the \So\ phase. Within a well defined lipid phase, the optical thickness of the bilayer is known, and we can convert the integrated qDIC phase to a radius, as discussed before. The right axis on \Fig{fig:PhaseAreaVsTemp} shows the calculated radius over the phase transition assuming \Ld\ phase parameters (using \So\ phase parameters yields 21\% smaller radius estimates). 

Tubes remaining connected to the bilayer stack, such as the one shown in \Fig{fig:PhaseAreaVsTemp}b, show a consistent behaviour in the evolution of their integrated optical phase during cooling. The integrated qDIC phase reduces slowly until the main phase transition begins in the planar bilayer regions, then drops more sharply reaching a minimum in the centre of the phase transition temperature range. This is followed by a gradual recovery that levels off once the transition is complete. Tubes that disconnect during the transition, such as the one shown in \Fig{fig:PhaseAreaVsTemp}a, represented by the pink line in \Fig{fig:PhaseAreaVsTemp}, do not follow this behaviour. This indicates that the consistent behaviour is driven by the common tension of structures connected to the planar bilayers during the phase transition discussed earlier. 

The mean integrated phase of the tubes that remain connected to the planar bilayers is shown as grey squares in \Fig{fig:PhaseAreaVsTemp}. In the \Ld\ phase, the phase area is approximately 7.8\,rad\,nm, which corresponds to a radius of 266\,nm. The radius then gradually decreases to a minimum around 107\,nm as the phase transition proceeds. After the transition, an integrated phase of around 5.2\,rad\,nm is measured, corresponding to a radius of approximately 141\,nm in the \So\ phase.

To discuss these changes in mean radius over the phase transition, we consider \Eq{eq:RadiusEq}. 
The bending rigidity $\eta$ changes the radius is proportional to $\sqrt{\eta}$, and has a minimum at $T_{\rm{m}}$, due to the known softening near first-order phase transitions. Indeed, experiments on the structurally similar lipid DMPC have shown that $\eta$ is lower by almost an order of magnitude at $T_{\rm{m}}$ relative to the \Ld\ phase\cite{DimovaACIS14},  which would lead to a three times smaller radius, consistent with the observed change between the equilibrium \Ld\ phase and the minimum during phase transition. 

However, $\eta$ in the \So\ phase is about five times the one of the \Ld\ phase,\cite{DimovaACIS14} therefore the radius would be expected to increase upon further cooling to about twice the one in the \Ld\ phase, in contrast to our observation of a 30\% radius reduction. This points to the influence of the membrane tension, $\sigma$, changing the radius proportional to $1/\sqrt{\sigma}$. The about three times lower radius after cooling into the \So\ phase than expected from the change of $\eta$ could be explained by an about 9 times higher $\sigma$.  

We recall that the tubes showing this characteristic shape evolution are contiguous with the unilamellar surface attached bilayer regions. As previously discussed, our understanding of the force driving tube formation is that it is more energetically favourable to pull lipids out of bilamellar regions of the film than to reduce the surface coverage of the unilamellar regions, due to the high strength of the interaction between the lower leaflet of the bilayer and the substrate. The higher lipid packing density in the \So\ phase drives the unilamellar regions to pull lipids from elsewhere in the system to maintain their surface area upon cooling, yielding the increase of $\sigma$ required to explain the observations. This consistent with the behaviour seen in \Fig{fig:TubeFormation}, also requiring an increase of $\sigma$ due to the surface adhered bilayers and the reduction of lipid bilayer area during the phase transition.

A kinetic effect could also be at play, assuming that in the \So\ phase the structure is frozen in place, not allowing lipid diffusion on the timescales observed. If the bilayer is rather described by a solid than a liquid on the relevant timescales, \Eq{eq:RadiusEq} not applicable. The radius would then be smaller than expected in the \So\ phase simply because no further lipid can be transported to the tube after freezing. However, the equilibration of the fluorophore after phase transition seen in \Fig{S-fig:TubeRadii} indicates diffusion over a few microns over the 24 minute observation time below the phase transition, which would allow sufficient transport, not supporting this hypothesis.

\begin{figure}
	\includegraphics[width=\columnwidth]{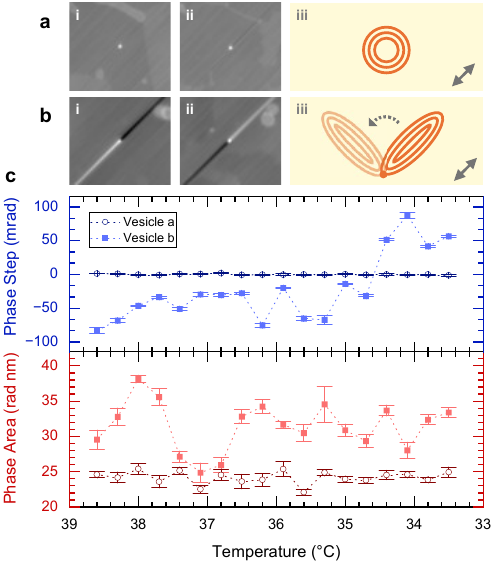}
	\caption{a,b) Small vesicles of \ac{dcpc} shown as \ac{qDIC} phase images (16\,\textmu m width) at 38.6\,\textdegree C  (i) and later after cooling to 33.5\,\textdegree C (ii), and top-down sketches of possible shapes and orientations with the DIC shear indicated by the grey arrows (iii). c) Fitted phase step ($2af$) and phase area ($2ac$), with the vesicle in a) represented by the open circles, and b) represented by the solid squares.}
	\label{fig:VesicleBirefringence}
\end{figure}

Notably, beyond tubular networks and bilayer edges, other membrane structures close to or below the diffraction limit were observed. For example, \Fig{fig:VesicleBirefringence} shows two \ac{dcpc} vesicles which, based on the magnitude of their phase profiles, are expected to be multilamellar (roughly 10 bilayers, assuming a 100\,nm radius). Both vesicles have similar integrated phases, but have phase steps, due to the birefringence, that differ by two orders of magnitude. From this, we can infer that the vesicle in \Fig{fig:VesicleBirefringence}a likely has a oblate profile lying flat, suppressing the effect of birefringence in qDIC, while the vesicle in \Fig{fig:VesicleBirefringence}b might be oblate standing on its side, creating strong birefringence effects. Further information was gained by changes in the birefringence over time, as the vesicles were gradually cooled to lower temperatures. While the phase step over the weakly birefringent vesicle remains essentially constant within error, the phase step of the more birefringent vesicle fluctuates considerably, and even changes sign (see \Fig{fig:VesicleBirefringence}c). The latter effect requires the vesicle to change its orientation reversing the direction of the birefringence, as illustrated in \Fig{fig:VesicleBirefringence}b,iii. 

\section{Summary and Outlook}

We have shown that lipid films with controlled lipid composition, spin-coated on a solid support, produce tubular structures, and we have quantified their radii by quantitative DIC microscopy, which is sensitive to nanometre-sized lipid thicknesses below the optical diffraction limit. 
Tubes were analysed in their equilibrium state using DOPC lipids that are in the liquid disordered (\Ld) phase at room temperature, and during formation with \ac{dcpc} lipids which transition to a solid ordered (\So) phase upon cooling. Tubular networks of DOPC lipids on glass showed radii of 150 - 200\,nm. Moreover, bilayers attached directly on glass have localised adhesion points, resulting in the formation of tubular networks, while tubes on top of lipid bilayers have little interaction with the surface and show extended straight geometries.

A complex arrangement of regions with different lamellarity is found, with surface tension created by the interaction with the glass surface dictating the tube diameter for the lipids in the liquid phase \Ld. Notably, owing to its sensitivity to birefringence, qDIC allowed us to distinguish between different tubular arrangements (e.g. concentric multilamellar versus twin unilamellar) and also reveals configurations such as folded bilayer edges, triple junctions and multilamellar nanovesicles.  

Tube formation on glass is found to be driven by a depletion of lipid material from bilamellar regions of the lipid film, and a mechanism is proposed whereby lipid depletion is due to migration into unilamellar patches that are strongly adhered to the glass surface. Using \ac{dcpc} lipid having a phase transition from \Ld\ to \So\ phase, the characteristic sizes of tubes during cooling was studied, created by the combination of changing surface tension as the lipid density changes and a changing bending rigidity, having a minimum during phase transition.  

The ability to produce well-defined supported lipid tubular structures with controlled composition and geometries using this model system opens up novel experimental possibilities for dissecting lipid-lipid interactions and biophysical properties of many biological filaments existing in nature.

\begin{acknowledgement}
	
	The microscope setup used was developed within the UK BBSRC Research Council (grant BB/H006575/1). D.R. acknowledges funding by an EPSRC DTA studentship, and by the UK BBSRC Research Council (grant BB/R021899/1). We acknowledge Joseph Bleddyn Williams for support in data analysis. Support by the Leverhulme Trust (grant RPG-2024-2019) is also acknowledged.  The authors thank Iestyn Pope for supporting the microscope setup and assistance in the data acquisition. 
	
\end{acknowledgement}



\providecommand{\latin}[1]{#1}
\makeatletter
\providecommand{\doi}
{\begingroup\let\do\@makeother\dospecials
	\catcode`\{=1 \catcode`\}=2 \doi@aux}
\providecommand{\doi@aux}[1]{\endgroup\texttt{#1}}
\makeatother
\providecommand*\mcitethebibliography{\thebibliography}
\csname @ifundefined\endcsname{endmcitethebibliography}
{\let\endmcitethebibliography\endthebibliography}{}

\end{document}